\documentclass[preprint,12pt]{elsarticle}

\usepackage{amssymb}
\usepackage{hyperref}
\hypersetup{
	colorlinks=true,
	linkcolor=blue,
	filecolor=magenta,      
	urlcolor=cyan,
}
\journal{International Journal of Hydrogen Energy}
\begin{document}

\begin{frontmatter}


 \title{Large-scale flame structures in ultra-lean hydrogen-air mixtures}
 \author[label1]{I.S. Yakovenko \corref{cor1}}
 \ead{yakovenko.ivan@bk.ru}
 \author[label1]{M.F. Ivanov}
 \author[label1]{A.D. Kiverin}
 \author[label1]{K.S. Melnikova}

 \address[label1]{Joint Institute for High Temperatures, Izhorskaya st. 13 Bd.2, Moscow, 125412, Russia}

 \cortext[cor1]{Corresponding author. Joint Institute for High Temperatures, Izhorskaya st. 13 Bd.2, Moscow, 125412, Russia. Tel.: +7 4954844433}

\begin{abstract}
The paper discusses the peculiarities of flame propagation in the ultra-lean hydrogen-air mixture. Numerical analysis of the problem shows the possibility of the stable self-sustained flame ball existence in unconfined space on sufficiently large spatial scales. The structure of the flame ball is determined by the convection processes related to the hot products rising in the terrestrial gravity field. It is shown that the structure of the flame ball corresponds to the axisymmetric structures of the gaseous bubble in the liquid. In addition to the stable flame core, there are satellite burning kernels separated from the original flameball and developing inside the thermal wake behind the propagating flame ball. The effective area of burning expands with time due to flame ball and satellite kernels development. Both stable flame ball existence in the ultra-lean mixture and increase in the burning area indicate the possibility of transition to rapid deflagrative combustion as soon as the flame ball enters the region filled with hydrogen-air mixture of the richer composition. Such a scenario is intrinsic to the natural spatial distribution of hydrogen in the conditions of terrestrial gravity and therefore it is crucial to take it into account in elaborating risk assessments techniques and prevention measures.

\end{abstract}


\end{frontmatter}

%%\begin{linenumbers}

\section{Introduction}
\label{sec1}

For decades combustion processes in lean gaseous mixtures have been a subject of study for numerous research groups worldwide. Apart from fundamental problems of the combustion theory, topics related to the combustion of the fuel-lean mixtures arise due to a wide diversity of its applications for many branches of industry and energy sector. Low emission of nitric oxides and other greenhouse gases defines the prospects of the propulsion devices and power plants that utilize lean combustion processes \cite{Evans200895}. Moreover, the detailed study of the lean flames features is of great concern when solving the issues related to fire and explosion safety \cite{Coward1952}. In particular, problems of assessment and minimization of potential risks of ignition of the combustible mixtures formed as a result of abnormal fuel emissions into surrounding atmosphere are of paramount interest \cite{Shapiro1957}. Studying accidents, which develop according to this scenario, special attention should be paid to combustion of the hydrogen-air mixtures. Ejection of hydrogen into the surrounding air is possible in various emergency situations, including accidents at the nuclear power plants \cite{international2011iaea}, where hydrogen is gradually accumulated in the atmosphere under containment as a result of the uncontrolled zircaloy oxidation process. At the same time, hydrogen is the perspective fuel, so the issues of its storage and transportation safety inevitably arise \cite{Venetsanos2003}. Depressurization of the hydrogen fuel tanks and pipelines is also one of the possibilities of the flammable mixture formation. In case of the hydrogen injection into the large confined spaces, formed combustible mixture composition could vary in a wide range. Herewith the hydrogen could be distributed non-uniformly in space \cite{Auban2007}, and locally the combustible mixture could have ultra-lean composition with equivalence ratio close to the lean flammability limit. Despite the low burning velocities and moderate pressure and thermal loads caused by such lean mixtures burn out, serious risks could be related with convective transfer of the lean flame into the location occupied with more chemically active composition (in case of natural vertical stratification of hydrogen \cite{Auban2007}) and its subsequent ignition, so more hazardous combustion processes could develop, such as fast deflagration or even detonation \cite{Alekseev2001}. All these considerations should be taken into account when formulating the risk mitigation measures and designing of the reliable explosion safety systems. 

In the hydrogen-air mixtures of near-stoichiometric and lean composition (with hydrogen content larger than ~10\%) the combustion propagates in a classic deflagration regime, that is widely discussed in a number of papers \cite{Jimenez2015,Yu2017}. As the lean flammability limit is approached, at hydrogen content less than 10\%, the main physical mechanism of the flame propagation switches from thermal diffusion to deficient specie diffusion into the reaction zone. Ya. B. Zel'dovich first theoretically predicted the possibility of the purely diffusive spherical flames formation in the ultra-lean mixtures or the so-called flame balls \cite{Zeldovich}. He provided theoretical analysis which implies that these structures are intrinsically unstable. However further theoretical and experimental studies shown that external forces such as gravity \cite{Brailovsky1997, Patnaik1991} or heat losses due to radiation \cite{Buckmaster1991, Ronney1990} or heat transfer to the cold walls \cite{Buckmaster1993} in microgravity conditions can stabilize ultra-lean flames. 

Flame propagating in the ultra-lean mixtures with Lewis number lower than unity under terrestrial gravity conditions is subjected to various instabilities such as hydrodynamic instability (Darrieus-Landau instability), thermo-diffusive instability proposed by Zel’dovich, Barenblatt and Sivashinsky and buoyancy-induced convective instability, which develops according to the general Rayleigh-Taylor mechanism. It is rather difficult to distinguish the influence caused by one or another instability on the flame propagation dynamics in the practically important environments that are characterized by complex geometries and terrestrial gravity force. One of the common ways to simplify the analysis is to conduct studies in microgravity or zero-gravity conditions, where the effects related with natural convection are negligible and the main role belongs to the thermo-diffusive instability development. A large number of theoretical and experimental studies in microgravity allowed to determine many significant features of the near limit lean flames. As it turned out, for mixtures with $Le < 1$ flame curvature intensifies combustion \cite{Ronney1990}. This feature of the lean flame is crucial for its stable propagation.Also, flammability limits and flame propagation regimes under the influence of thermo-diffusive instability were thoroughly examined at microgravity conditions. However, these results are not sufficient for solving practical fire and explosion safety issues at terrestrial gravity conditions. Convective transport of the flame and gas-dynamical flows induced by the flame motion should be taken into account as they can alter flammability limits, flame propagation velocities and affect the observed flame structure. Theoretical studies have shown that convection-induced instabilities have a crucial impact on the slow ultra lean flames and stabilize flame structure during flame upward propagation \cite{McIntosh1985,Pelce1982}. Further numerical simulations proved theoretical estimations \cite{Patnaik1991, PatnaikOran1991} for lean H$_2$-O$_2$-N$_2$ mixtures and reproduced initial stages of the flame propagation out from the ignition source and formation of the cap-shaped flame structures observed in experiments \cite{Levy1965,Sun2014}. 

It is useful to note that ultra-lean flame is subjected to the gas-dynamical factors related with expansion of hot combustion products in different manner than widely studied freely propagating deflagrative flames \cite{Liberman2004, Pan2008}. First of all, lower reactivity of the ultra-lean compound does not provide sufficient rate of heat release for stable deflagration wave formation. Due to this fact expansion factor is not enough for stable outwardly flame propagation. In case of microgravity a stable flame ball is formed supported by the diffusion mechanisms discussed above. In case of terrestrial conditions the main upward direction of flame propagation establishes due to buoyancy of hot products. In addition to this the low value of burning rate does not provide intensive rise of intrinsic Landau-Darrieus instability \cite{Liberman2004,Gostintsev1988} at the background of outward flame propagation in the terrestrial gravity field.

However, discussed studies were mainly aimed on fundamental analysis of various features of ultra-lean flames and were abstracted from the issues related to fire or explosion safety. In particular, a detailed analysis of the combustion dynamics and its stability mechanisms in the real-scale domain is necessary to prevent accidents caused by the convective transfer of the flame in the lean mixture. A significant effort in this direction was made in \cite{Hernandez-Perez2015,Shoshin2011,Zhou2017}, where H$_2$-CH$_4$-air lean flames were stabilized in the cylindrical burner and analyzed both experimentally and numerically. Important results on ultra-lean flame structure, its stability, and issues related to its numerical modeling were obtained by the authors. However, dynamics of freely propagating ultra-lean flames affected by natural convection in large-scale domains is still not thoroughly examined. In this study, we present novel results on numerical simulation of the non-stationary flame propagation in the large-scale domain filled with the ultra-lean hydrogen-air mixture that can be used for solving real issues in a field of hydrogen safety. 

\section{Problem setup and numerical method}
\label{sec2}

\begin{figure}[h]
	\centering\includegraphics[width=0.75\linewidth]{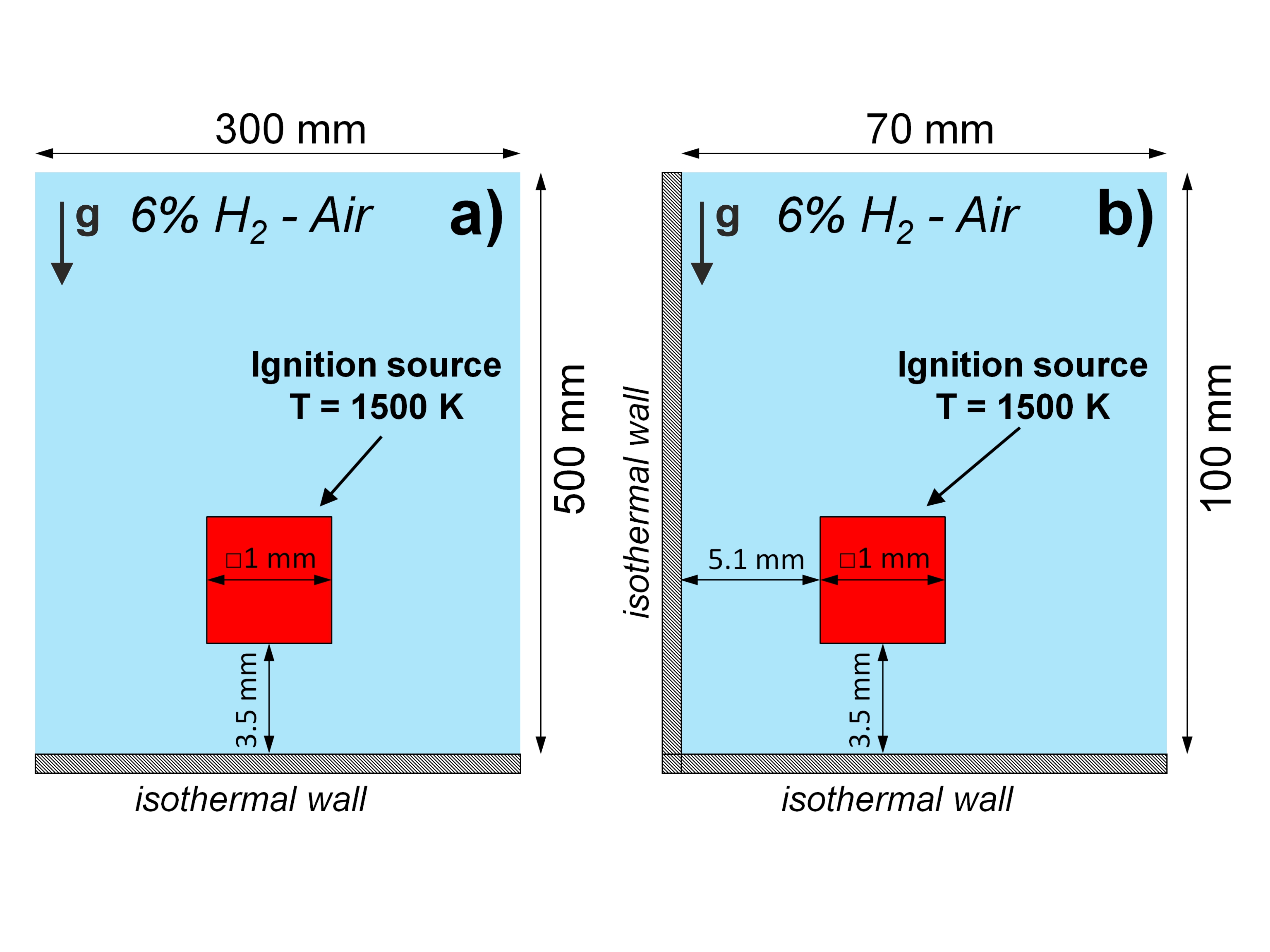}
	\caption{\label{figure1} Computational domain setup for a -- flame propagating out from the wall, b -- flame propagating out from the corner.}
\end{figure}

In this paper, we solved numerically the problem of flame ball formation and propagation in the semi-unconfined space filled with the ultra-lean hydrogen-air mixture with 6\% hydrogen content at normal initial conditions ($T=300$ K, $p=1$ atm). Numerical domain represented half-space confined with the wall from the bottom and outlet conditions on the top and both sides. Bottom wall temperature was equal to the ambient temperature of the mixture ($T_{wall} = 300$ K). The ignition of the mixture was modeled with the instantaneous energy input into the small region corresponded to the local mixture heating up to 1500 K. Schematically the problem setup is shown in Figure \ref{figure1}. In addition, to understand the geometry effect on the stability of ultra-lean flame ball we solved the problem with ignition source located in the vicinity of the vertical wall (Fig. \ref{figure1}b). Vertical wall temperature in that case was also set equal to $300$ K. Cartesian grid spacing was $0.2$ $mm$. That value was determined from one-dimensional flame velocity convergence tests for the considered combustible mixture composition.

To reproduce the dynamics of flame ball propagating through the ultra-lean hydrogen-air mixture under the action of buoyancy force the Navier-Stokes system of conservation laws was used in the low Mach number approximation. To reproduce the features of the simulated processes related with combustion the thermal conductivity, multicomponent diffusion and energy release associated with chemical transformations were taken into account. Chemical kinetics was calculated according to the contemporary kinetic scheme presented in \cite{Keromnes2013}. Governing equations are the following:

\[\frac{{\partial \rho }}{{\partial t}} + \nabla \left( {\rho \vec u} \right) = 0\]

\[\frac{{\partial \rho {Y_k}}}{{\partial t}} + \nabla \left( {\rho {Y_k}\vec u} \right) = \nabla \left( {\rho {Y_k} \vec {V_{k}}} \right) + \rho {\left( {\frac{{d{Y_k}}}{{dt}}} \right)_{chem}}\]

\[\frac{{\partial \rho {u_i}}}{{\partial t}} + \frac{{\partial \rho {u_i}{u_j}}}{{\partial {x_j}}} =  - \frac{{\partial p'}}{{\partial {x_i}}} + \frac{{\partial {\sigma _{ij}}}}{{\partial {x_j}}}\]

\begin{eqnarray*}
	\frac{{\partial \rho \varepsilon }}{{\partial t}} + \nabla \left( {\rho \varepsilon \vec u} \right) =  - {p_0}div\left( {\vec u} \right) + \nabla \left( {\kappa \nabla T} \right) \\
	+ \nabla \left( \sum\limits_k{\rho {h_k Y_k \vec V_{k}}} \right) + \sum\limits_k {\rho {h_k}{{\left( {\frac{{d{Y_k}}}{{dt}}} \right)}_{chem}}} 
\end{eqnarray*}

\[{p_0} = \left( {\gamma  - 1} \right)\rho \varepsilon \]

\[d(\rho \varepsilon) = C_V(T)dT\]

Here $\rho$ is the mass density, $\vec u$ is the mass velocity, $Y_k$ is the molar fraction of $k$-th component of gaseous mixture, $p'$ - dynamic component of pressure fluctuations, which is by the order of magnitude much smaller compare with thermodynamic pressure $p_0$, which assumed to be constant in the process of flame propagation,  $\sigma _{ij}$ - components of viscous stresses tensor, $\varepsilon$ - specific internal energy, $T$ - temperature, $\kappa$ - thermal conductivity coefficient, $h_k$ enthalpy of formation of $k$-th component of gaseous mixture, $\vec V_{k}$ - diffusion velocity vector of $k$-th component, $C_V(T)$ - specific heat capacity at constant volume, $\gamma$ - ratio of heat capacities at constant pressure and constant volume. Term $\left(\frac{d{Y_k}}{dt} \right)_{chem}$  represents the change in molar fraction of $k$-th component due to the chemical reactions.

Specific heat capacities and enthalpies of formation are calculated according to the JANAF tables \cite{Chase1998}. To determine dynamic pressure a Poisson equation is solving at each time step. Herewith the value of velocity divergence is calculated from the equation for internal energy taking into account, on the one hand, the energy change due to heat release and thermal conductivity and on the other hand the energy change according to the equation of state. Such an approach allowed us to utilize explicit numerical scheme distinct to the more conventional approach \cite{Tomboulides1997} where correction is necessary if temperature-dependent specific heat capacities are used.

Diffusion model is based on the zeroth-order Hirshfelder-Curtiss approximation \cite{HirschfelderCurtissBird1964}. Mixture averaged transport coefficients were obtained from the first principles of the gas kinetics theory \cite{KeeColtrinGlarborg2003}. Diffusion velocities were calculated taking into account correction velocity approach proposed in \cite{Coffee1981}. Thermal diffusion or Soret effect was not incorporated in the used model. Despite its considerable influence on the structure and burning velocities of freely propagating flames in lean hydrogen-based mixtures rigorously studied in \cite{Ern1998,Ern1999,Grcar2009}, here we aimed to provide a qualitative description of the combustion wave development, without addressing issues of precise reproduction of the flammability limits and burning characteristics. Although thermal radiation represents one of the stabilizing factors for near-limit flames, in considered problem setup of flame propagation in the closed vessel with isothermal walls under normal gravity conditions, thermal losses on the cold walls and convection are more significant processes for flame stabilization than thermal radiation \cite{Zhou2017}. Thereby thermal radiation is also neglected in the utilized physical model. 

Governing equations were solved using the explicit euler-lagrangian numerical procedure that was previously successfully applied to solve a wide range of problems of the hydrogen-based gaseous mixtures non-steady combustion processes \cite{Ivanov2015, Kiverin2016, Ivanov2017}. The system of ODEs defining the combustion kinetics was solved by Gear method.

\section{Results and discussion}
\label{sec3}

\begin{figure}[h]
	\centering\includegraphics[width=0.75\linewidth]{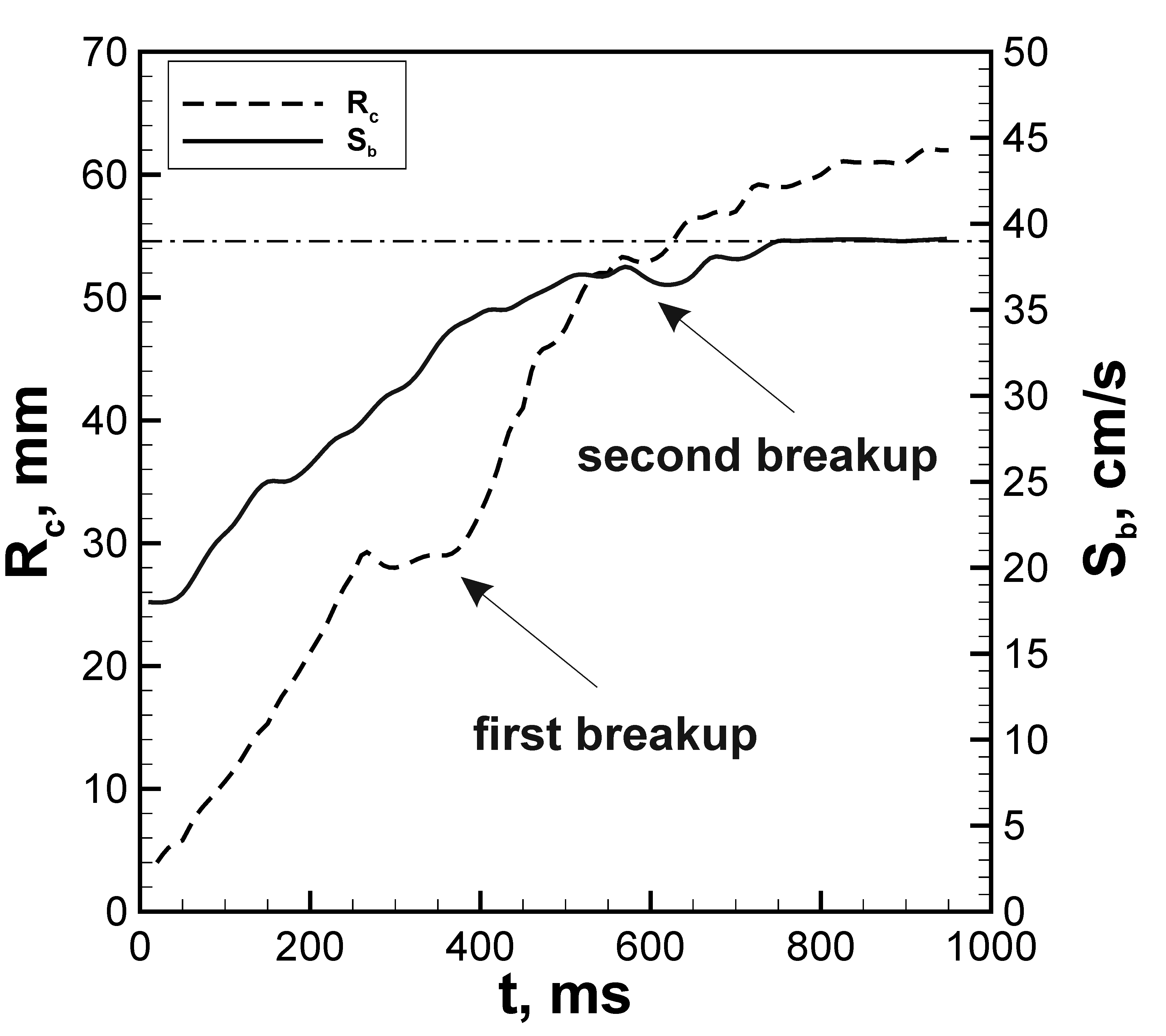}
	\caption{\label{figure2} Histories of the flame front leading point velocity ($S_b$) and curvature radius ($R_c$) of the main flame core. Theoretically predicted terminal velocity for the closed bubble with curvature radius 62 $mm$ is $38.9$ $cm/s$ and is shown by dash-dotted horizontal line.}
\end{figure}

\begin{figure*}[h]
	\centering\includegraphics[width=1.0\linewidth]{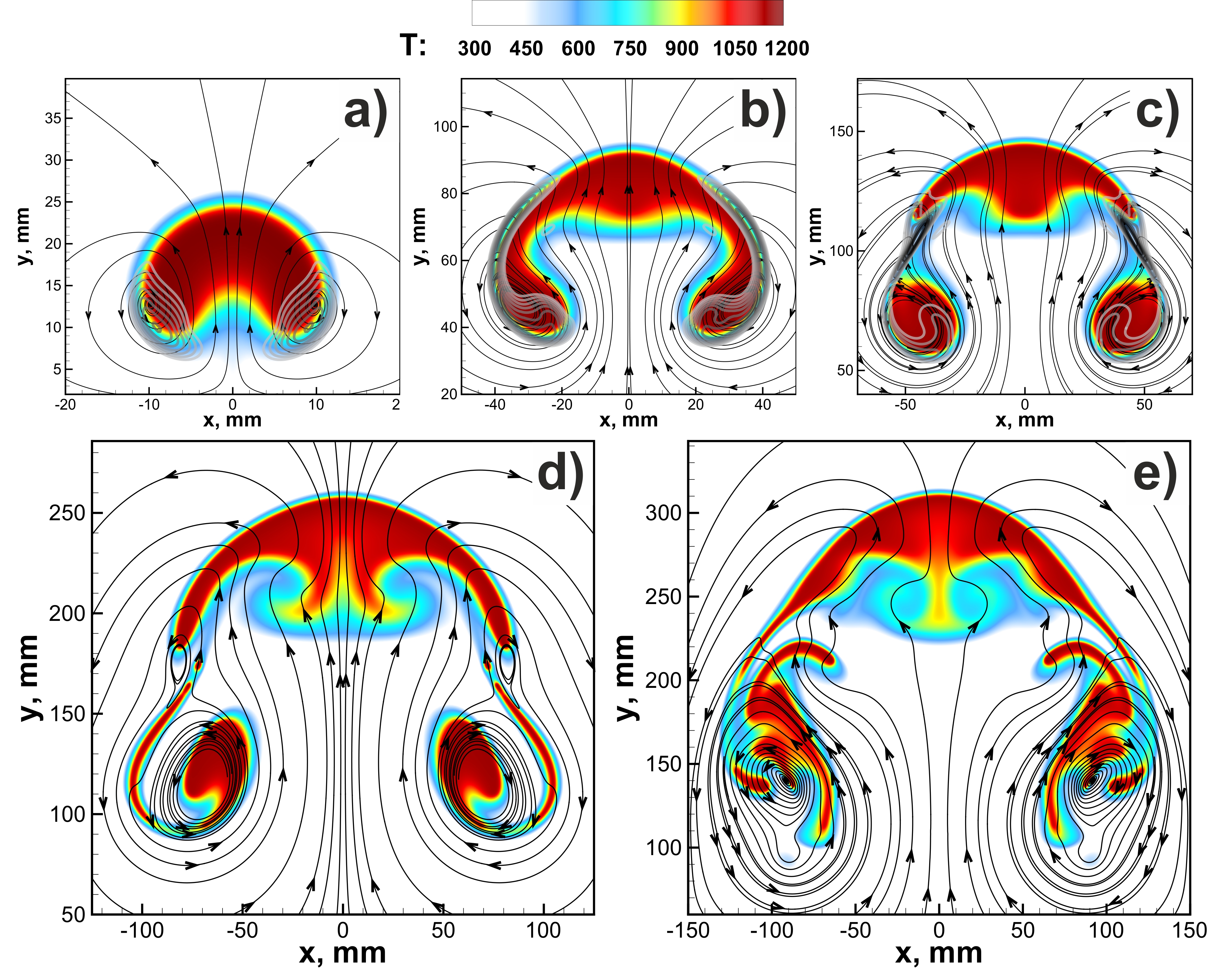}
	\caption{\label{figure3} Ultra-lean hydrogen flame evolution. a -- 100 $ms$, b -- 350 $ms$, c -- 500 $ms$, d -- 800 $ms$, e -- 950 $ms$. Background is colored by temperature. Streamlines are illustrated by the solid lines with arrows. On figures a, b, c vorticity contours colored by the vorticity magnitude (from light to dark) are also presented.}
\end{figure*}

Let us consider in details the stages of flame ball evolution in the ultra-lean hydrogen-air mixture. Immediately after ignition spherical flame kernel is formed in the region of the ignition source. On this early stage combustion process is governed purely by the diffusion of the deficient hydrogen into the reaction zone, so the flame kernel represents a classic flame ball similar to the case of micro-gravity. According to the theoretical analysis of the similar setup by \cite{Buckmaster1993}, the flame stability at this early stage is mainly supported by heat losses to the wall. As the combustion proceeds, the flame kernel is expanding beyond the critical size and hot combustion products become subjected to buoyancy force \cite{Leblanc2012}. From now on the main mechanism of the flame propagation is the convective rising in the gravitational field. As the flame front is a density discontinuity, its motion can be addressed qualitatively by the theory of the closed rising bubble \cite{Bychkov2000}.The early stage of the flame upward propagation is characterized by the nearly constant acceleration of the flame leading point, which is the highest point of the flame front (see. Fig. \ref{figure2}). Gas-dynamical flows induced by the flame kernel rising are also resembling those develop during the bubble rising \cite{Hua2008}. Recirculation zone consisting of two large vortices is formed in the wake of the flame directly after its upward moving starts \cite{Higuera2014} (see streamlines presented in figure \ref{figure3}a). Fresh mixture is attracted to the trailing edge of the flame kernel by the vortical flows and deficient reactant diffusion that determines non-zero reaction rate on the flame trailing edge \cite{Hernandez-Perez2015}. 

\begin{figure*}
	\centering\includegraphics[width=1.0\linewidth]{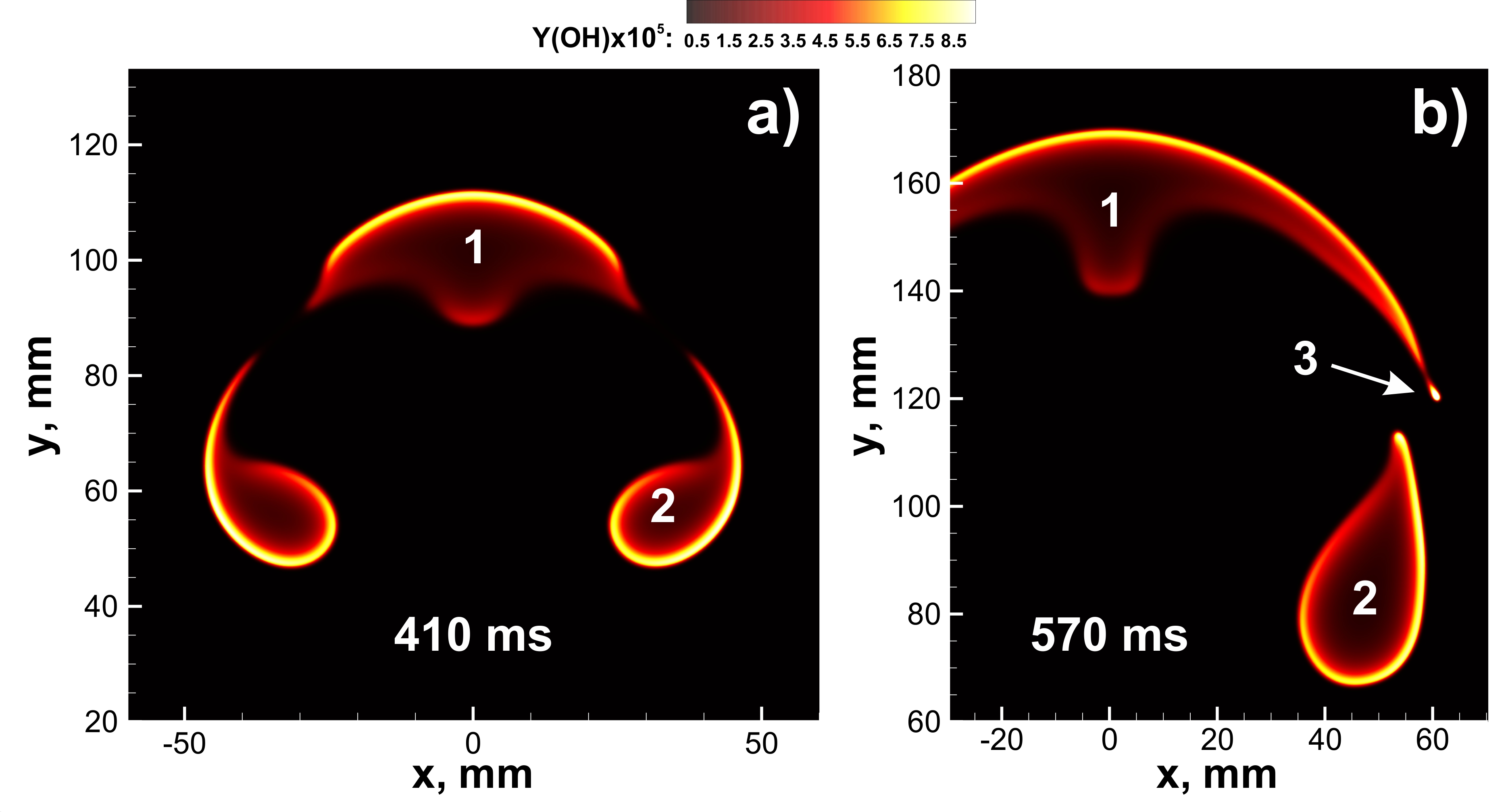}
	\caption{\label{figure4} Flow patterns in the process of flameball breakup. 1 -- main flame core. 2 -- first separated burning kernel (stable). 3 -- second separated burning kernel (unstable).}
\end{figure*}

Flame stretch due to vortical flows leads to the characteristic cap-shaped flame structure formation (see Fig. \ref{figure3}b), which was experimentally and numerically observed for both bubbles \cite{Hua2008} and near limit flames \cite{Leblanc2012,Sun2014,Zhou2017}. Similarly as for the closed bubble with small values of the surface tension \cite{Tripathi2015} cap-shaped structure of the flame core is not stable. Developing of the high strain rates on the side surface of the flame due to vortical motion leads to the peripheral breakup of the flame core with the formation of two satellite flame kernels on the sides of the main flame core (see vorticity contours on Figs. \ref{figure3}b,c). Flame breakup preserves combustion processes in the satellite kernels. In contrast to the bubble dynamics, where satellite bubbles do not change in size, satellite flame kernel represents an active reaction zone and becomes larger consuming fresh combustible mixture. After the breakup, the acceleration of flame core rising  sharply decreases. Linear stage that is observed from the beginning until approximately 400 $ms$ is replaced by the non-linear stage, which is characterized by multiple peripheral breakup events. It is known that closed bubble terminal rising velocity is proportional to the $\sqrt{R}$, where $R$ is the curvature radius near the bubble symmetry axis \cite{Davies1950}. Thus the rising velocity temporal evolution could be interpreted in terms of the curvature radius. Time dependence of the flame core curvature radius near the symmetry axis is presented on figure \ref{figure2}. The figure shows that after the flame breakup the curvature radius becomes almost constant for a short period of time, leading to the rising velocity increase temporary slowdown. After the breakup, flame core represents a lens-shaped flame ball which evolves similarly to the flame ball at early stages, transforming into the cap-shaped flame due to the combustion at the trailing edge of the side surface of the flame ball. This process is amply demonstrated in figure \ref{figure4} where the OH field illustrates the structure of reaction zone inside main flame core and satellite kernels. One can vividly observe the intensification of combustion at the trailing edge and downward propagation of the flame at its side surface. Subsequently, this new cap-shaped flame core breaks up leading to the formation of a new pair of the kernels behind. It should be noted that this second breakup event (at time instant 570 $ms$) is less intensive compare with the first one (at 410 $ms$), so the new kernel occurs to be rather small and quenches fast. After second breakup leading point velocity becomes governed solely by the flame front curvature radius near the symmetry axis and can be approximated by $S_b \sim \sqrt{R_c}$ dependence.

\begin{figure*}
	\centering\includegraphics[width=1.0\linewidth]{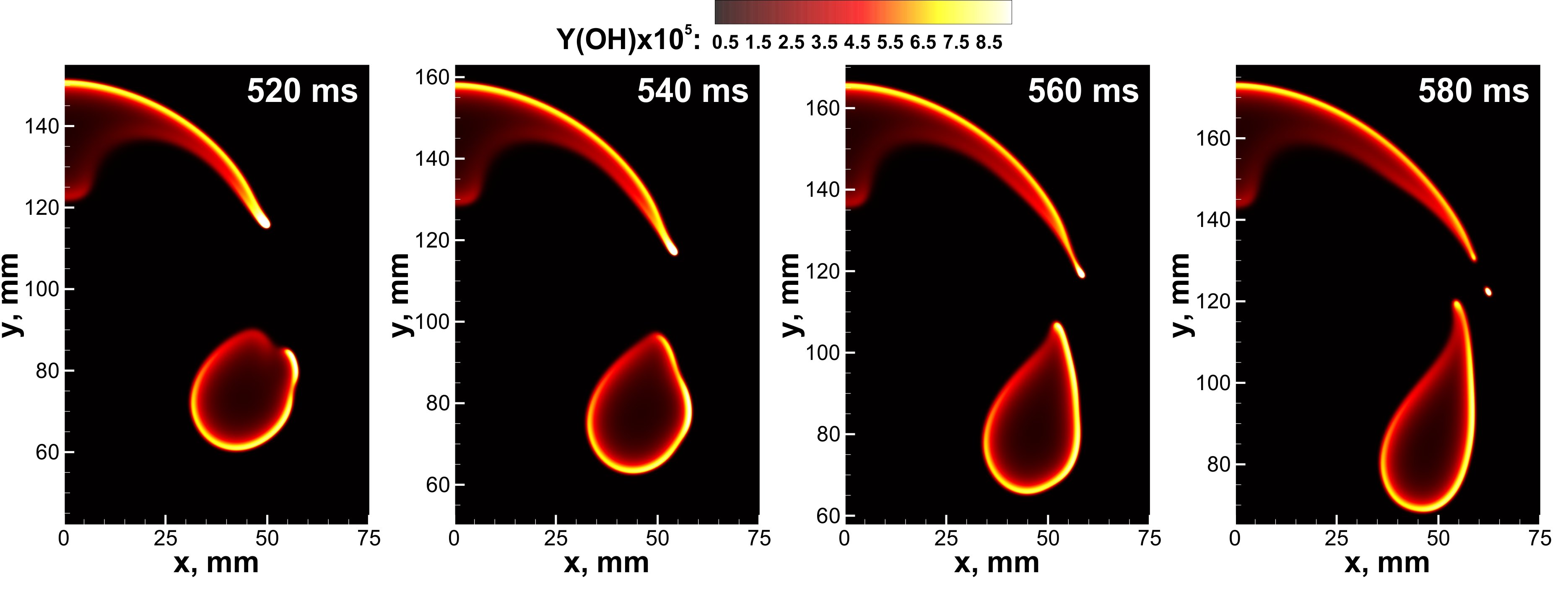}
	\caption{\label{figure5} Separated kernel upward propagation with acceleration in a thermal wake behind the main flame core.}
\end{figure*}

It is important to note that after the flame separation there are areas of high temperature between satellite kernels and the main flame core. Due to high local background temperature combustion intensifies in these areas that lead to the flame reassembly (see Fig. \ref{figure5}). Thus at time instant $\sim 520$ $ms$ the separated kernel, while propagating upwards after the main flame core, enters the heated region that provides local flame acceleration, and this kernel overtakes the trailing edge of the main flame core at time instant $\sim 650$ $ms$. The interaction between two flame kernels (the main one and the separated one) affects the structure and dynamics of the trailing surface of the main flame core. It should be, however, noted that at this time the main lens-shaped flame core is widely spread in space, so the changes in the shape of the trailing surface have almost no influence on the dynamics of the upper surface of the flame. Moreover, the next breakup of the side surface of the main flame core (at $\sim 920$ $ms$) also causes no change in the dynamics of the leading edge of the flame. And shortly after this one can observe the stagnation in the flame speed (Fig. \ref{figure2}), which is not anymore affected by the breakup events and interactions with separated burning kernels. By that time, main flame core attains lens-shaped structure with almost constant flame curvature radius near the leading point (Fig. \ref{figure2}). Such a behavior in the flame speed and curvature radius dynamics is similar to the bubble case. Obtained terminal velocity is in good agreement with theoretical estimation $S_b = 0.5 \sqrt{g R}$ , where $g$ is the gravity acceleration and $R$ is the bubble curvature radius \cite{Davies1950}. For the considered case of combustion in 6\% H$_2$-air mixture $S_b$ can be estimated as $\sim 38.9$ $cm/s$.

\begin{figure}[h]
	\centering\includegraphics[width=0.75\linewidth]{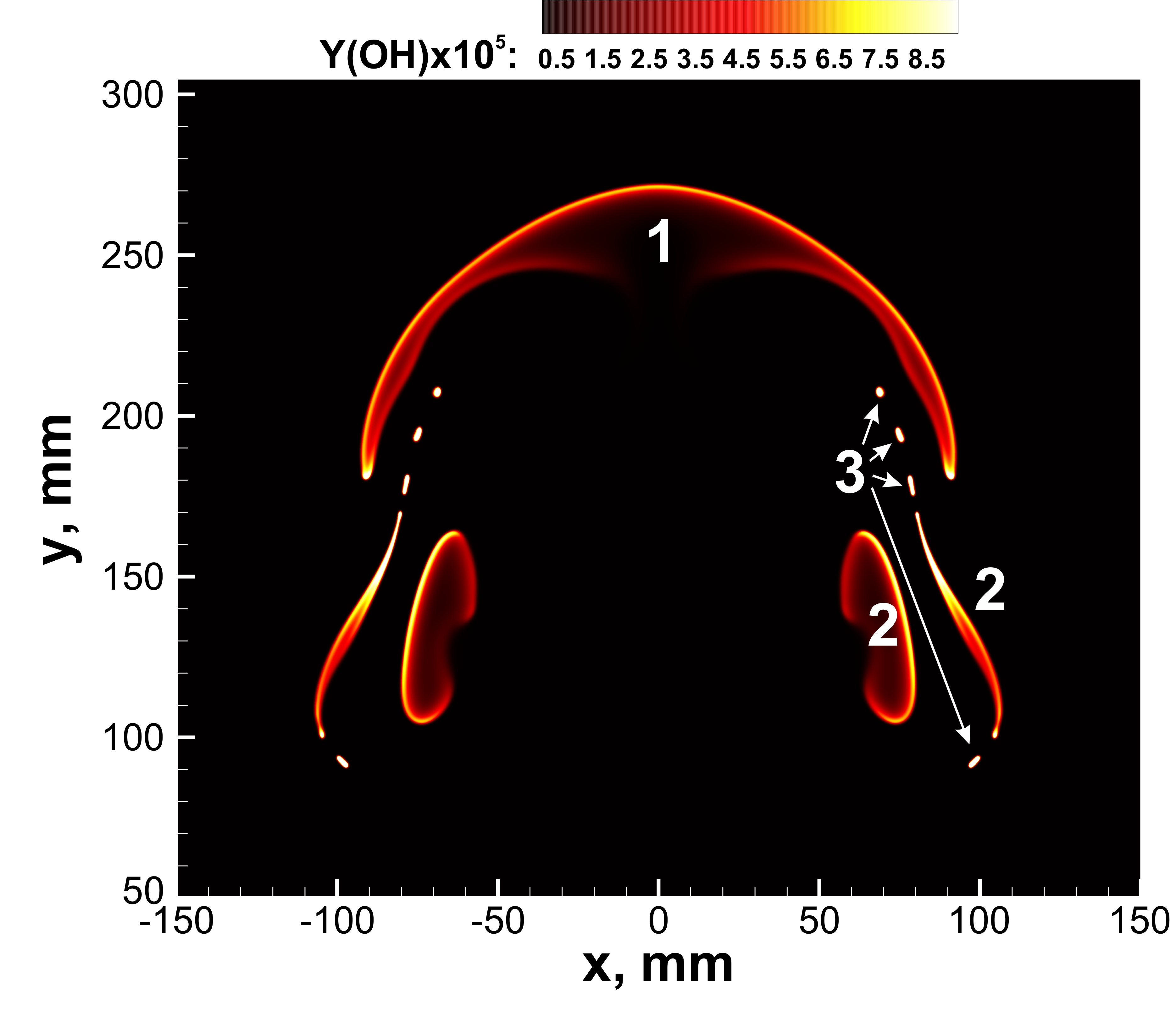}
	\caption{\label{figure6} Complex structure of the ultra-lean flame ball at 840 $ms$. 1 -- main flame core. 2 -- fragments of separated burning kernel. 3 -- burning kernels formed in the thermal wake.}
\end{figure}

Flame kernel previously reassembled with the main one has separated from the main core again at time instant $\sim 690 ms$. After this, it evolves as an independent burning kernel again. This kernel undergoes breaking up as well as the main flame core does, that leads to the formation of the complex structure of many independent satellite kernels widely spread in space (see Fig. \ref{figure6}). Satellite kernels are easily picked up by the vortical flows and drift away from the centerline of the combustion vessel, so the possible impact area expands forming a cone with apex point at the initial ignition region. As these satellite kernels propagate inside the thermal wake behind the main flame core there are events of their upward acceleration, so there is almost no lag between them and the main flame core. In such a way the following stable regime of the flame propagation is established. The main flame core representing a cap-shaped flame ball propagates with constant speed and followed by a widely spread flame "skirt" consisting of independent combustion zones. Such a stable flame structure could be treated as a possible way of energy transfer through the ultra-lean region. In case of natural hydrogen stratification in terrestrial conditions and energy input from the bottom, such an energy transfer could cause the ignition of richer and more chemically active hydrogen-air mixtures, in which the deflagrative or even detonative combustion could be initiated. In case of nearly uniform hydrogen distribution in space, such a process of energy transfer could damage electrical wiring or any other equipment resulting in technical system malfunction.

\begin{figure*}[h!]
	\centering\includegraphics[width=1.0\linewidth]{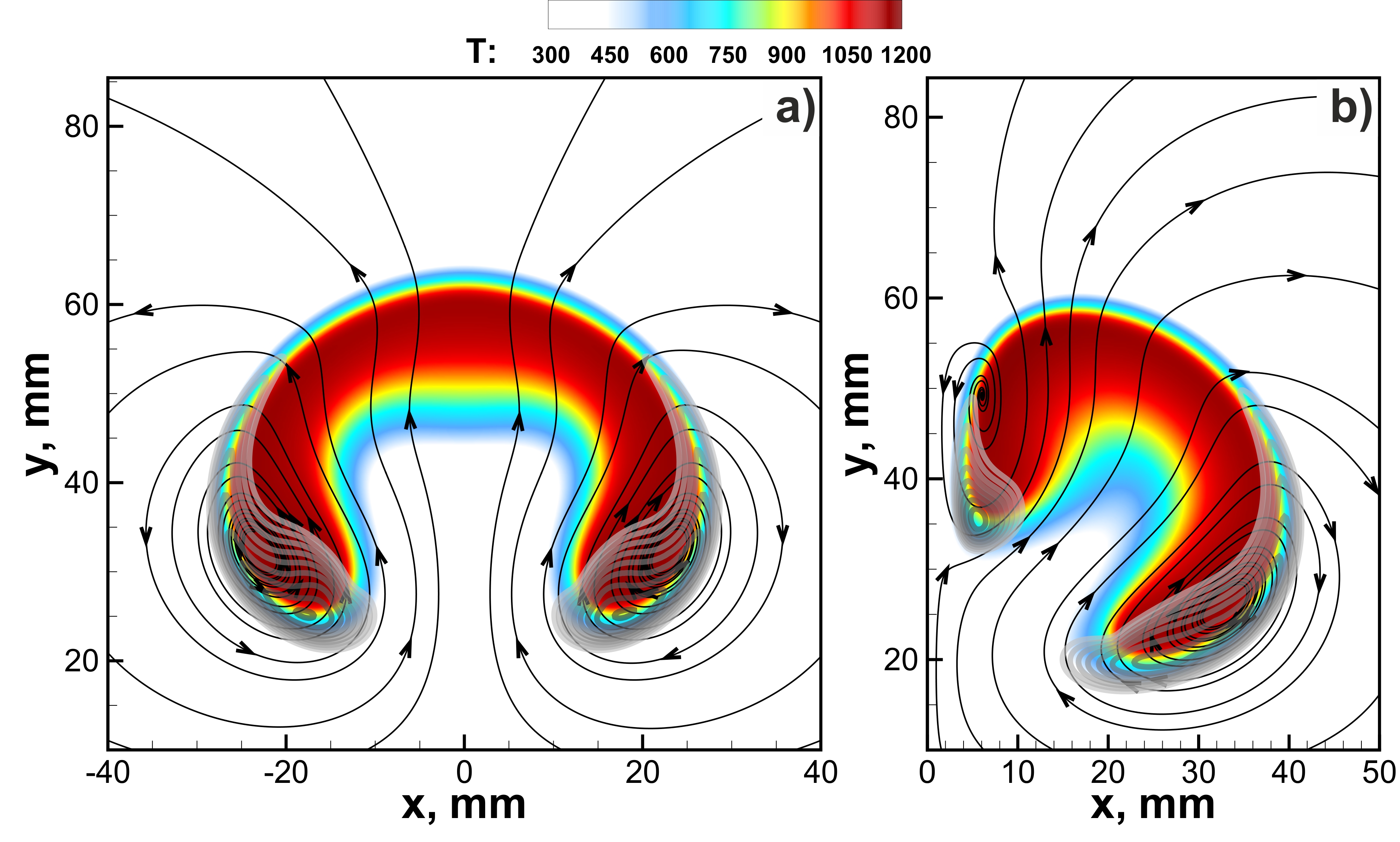}
	\caption{\label{figure7} Ultra-lean flame structures at 250 $ms$ for cases of freely propagating flame (a) and flame propagating along vertical wall (b). Background is colored by temperature. Streamlines are illustrated by the solid lines with arrows. Vorticity contours colored by the vorticity magnitude (from light to dark).}
\end{figure*}

Since the spatial evolution of the ultra-lean flame plays a great role in its dynamics it is reasonable to understand what limitations could be caused by the confinement. In this regard, we additionally studied the flame evolution in the same mixture but with the ignition source located near the vertical wall. The behavior of the flow near the wall surface determines the restriction of the main flame core spreading. Moreover, the direction of flame propagation slightly changes, and the flame propagates upward and little away from the wall. The breakup of the near-wall flame surface is also limited due to the heat and momentum losses in the boundary layer and therefore due to less intensity of combustion at trailing edge of the side surface of the flame. As a result, all the phenomena described above develops only at the side surface of the main flame core detached from the vertical wall. In two-dimensional problem setup, this causes an almost twice smaller spatial spread of the main flame core and the flame skirt. Due to this the flame speed occurs to be smaller ($S_b = 23$ $cm/s$ compare with $S_b = 39$ $cm/s$ in unconfined space) as well as the combustion area does (see Fig. \ref{figure7}, illustrating the flame ball structures for two cases at fixed time instant). Thus it seems to be useful to utilize the congestion of the volume filled with ultra-lean hydrogen-air mixture to prevent energy transfer via the stable mechanism considered above. For sure this can not be treated as a unified measure as soon as the richer mixture burning in the deflagrative regime could cause much more violent flame propagation regimes inside congested volumes. Therefore such a technique could be proposed only for the regions where the ultra-lean mixture composition can be predicted with a high probability.

\section{Conclusions}
\label{sec5}

The results of numerical analysis of ultra-lean hydrogen-air flame propagating in semi-unconfined space allowed to describe the peculiarities of the flame evolution. It is shown that the flame structure is fully determined by the convection flows related with flame ball rising in the terrestrial gravity conditions. The regime of flame ball propagation and its structure agree well with the theory elaborated for axisymmetric gaseous bubbles in the liquid. As well as in the case of the gaseous bubble, there is deformation of the initial flame ball, formation of a cap-shaped flame core and its breaking up at the trailing edge of the flame ball side surface. Unlike the case of the gaseous bubble, the flame structure is also defined by the combustion mechanisms. Thereby the separated burning kernels evolve inside the thermal wake behind the main flame core, that leads to the formation of stable flame skirt consisting of separated burning kernels propagating immediately behind the main flame core and spreading in space. Obtained structure of the ultra-lean hydrogen-air flame represents the stable cap-shaped main core followed by the widely spread in space multi-kernel combustion zone. This phenomenon is studied on relatively large spatial scales (compare e.g. with recent experiments \cite{Leblanc2012}) and occurs to be stable. Herewith the stability is dictated by the intrinsic gas-dynamical effects taking place in a reacting flow. This allows concluding that such structures represent a stable mechanism of energy transfer from the energy sources through the regions filled with ultra-lean hydrogen-air mixtures. In turn, this mechanism could cause the ignition of richer hydrogen-air mixtures located above energy sources in terrestrial conditions or cause the technical system malfunction due to heat effects. 

\section*{Acknowledgements}
The work was funded by the grant of Russian Science Foundation No. 14-50-00124. The research is carried out using the equipment of the shared research facilities of HPC computing resources at Lomonosov Moscow State University. We also acknowledge high-performance computing support from the Joint Supercomputer Center of the Russian Academy of Sciences.

%%\end{linenumbers}
\section*{References}

\bibliography{bibliography}


%% \begin{thebibliography}{00}


%%\bibitem{}

%%\end{thebibliography}
\end{document}